\documentclass[%
 reprint,
nofootinbib,
 amsmath,amssymb,
 aps,
]{revtex4-1}

\usepackage{graphicx}% Include figure files
\usepackage{dcolumn}% Align table columns on decimal point
\usepackage{bm}% bold math
\newcommand{\xv}{\boldsymbol{x}}
\newcommand{\Xv}{\boldsymbol{X}}
\newcommand{\Fv}{\boldsymbol{F}}
\newcommand{\Gv}{\boldsymbol{G}}
\newcommand{\Phiv}{\boldsymbol{\Phi}}
\newcommand{\erm}{\mathrm{\erm}}

\begin{document}

\preprint{APS/123-QED}

\title{Sparse Identification of Slow Timescale Dynamics}% Force 

\author{Jason J. Bramburger}
\email{jbramburger7@uvic.ca}
\affiliation{Department of Mathematics and Statistics, University of Victoria, Victoria, BC, Canada. V8W 2Y2}
\author{Daniel Dylewsky}
\email{dylewsky@uw.edu}
\affiliation{Department of Physics, University of Washington, Seattle, WA, USA. 98195}
\author{J. Nathan Kutz}
\email{kutz@uw.edu}
\affiliation{Department of Applied Mathematics, University of Washington, Seattle, WA, USA. 98195}

\begin{abstract}
Multiscale phenomena that evolve on multiple distinct timescales are prevalent throughout the sciences.  It is often the case that the governing equations of the persistent and approximately periodic fast scales are prescribed, while the emergent slow scale evolution is unknown. Yet the course-grained, slow scale dynamics is often of greatest interest in practice. In this work we present an accurate and efficient method for extracting the slow timescale dynamics from signals exhibiting multiple timescales that are amenable to averaging. The method relies on tracking the signal at evenly-spaced intervals with length given by the period of the fast timescale, which is discovered using clustering techniques in conjunction with the dynamic mode decomposition. Sparse regression techniques are then used to discover a mapping which describes iterations from one data point to the next. We show that for sufficiently disparate timescales this discovered mapping can be used to discover the continuous-time slow dynamics, thus providing a novel tool for extracting dynamics on multiple timescales.  
\end{abstract}

\maketitle

%\tableofcontents

\section{\label{sec:Intro}Introduction}

Many physical phenomena exhibit multiscale dynamics where the fastest timescale is relatively simple to both observe and predict, while the emergent long timescale dynamics are unknown. Examples abound in physics and range from tidal amplitudes \cite{Wunsch,Wunsch2}, to molecular dynamics simulations~\cite{andersen,karplus}, to atmospheric dynamics \cite{Palus}, to the motion of the planets \cite{Brasser,Laskar,Malhotra}. Mathematically, systems which exhibit multiscale dynamics are expensive to simulate since the fastest scales must be accurately resolved, and when the fast timescale is approximately periodic, this expense is primarily used to simulate predictable dynamics. In systems where the scale separation is explicit, it is sometimes the case that one can average the system to obtain the leading-order dynamics of the slow timescale evolution \cite{Guckenheimer,Kifer,Sanders}, although many systems lack an obvious separation of scales or even governing equations, meaning that novel methods for extracting the slow timescale dynamics must be developed.  

In this work we present a computationally cheap and efficient method that integrates machine learning and multiscale modeling for extracting and forecasting slow timescale dynamics of a multiscale system. Unlike averaging techniques, the proposed mathematical architecture learns a nonlinear dynamical system characterizing the slow-scale behavior. The setting for the problem is that we are given a signal $\Phiv(t,\varepsilon t)$ in $\mathbb{R}^d$ which contains terms dependent on $t$ contributing to the fast dynamics and terms dependent on $\varepsilon t$, for some $0<\varepsilon \ll 1$, contributing to the slow dynamics. We assume that the fast timescale terms are periodic, so $\Phiv$ satisfies 
\begin{equation}\label{Signal}
	\Phiv(t+T,\varepsilon (t+T))=\Phiv(t,\varepsilon (t+T))\approx\Phiv(t,\varepsilon t),
\end{equation}
for all $t$ and some $T > 0$. The timescales of $\Phiv$ are assumed to be sufficiently disparate so that the slow dynamics are nearly constant on the intervals $[t,t+T]$, coming from the size of $\varepsilon$ relative to the fast period $T$. It is through these assumptions that we see that the signal exhibits a fast oscillation and a slow drift, making it amenable to averaging. 

Suppose the signal is given on some finite timescale and we wish to understand the physics governing the slow timescale evolution so that the signal can be reconstructed and forecast far beyond the given time window. Since the fast-scale dynamics are relatively simple we would like to `average' these dynamics out to forecast only the slow-scale variable. Knowing the fast-scale period $T>0$ naturally leads to tracking the signal after each period:
\begin{equation}\label{xns}
	\xv_n = \Phiv(0,\varepsilon nT), \quad n = 0,1,2,3,\dots
\end{equation}
since $\Phiv(nT,\cdot) = \Phiv(0,\cdot)$ by periodicity. It follows that understanding the signal at $t = nT$ requires an understanding of the slow timescale dynamics. Our goal is to discover a mapping $\Fv:\mathbb{R}^d \to \mathbb{R}^d$ so that 
\begin{equation}\label{SlowMap}
	\xv_{n+1} = \Fv(\xv_n),
\end{equation}
for all $n \geq 0$. Beyond this, if we suppose there exists a function $\Gv:\mathbb{R}^d \to \mathbb{R}^d$ such that the slow timescale dynamics of a signal satisfying Eq.~\eqref{Signal} are governed by the ordinary differential equation (ODE) 
\begin{equation}
    \partial_2\Phiv(0,t) = \Gv(\Phiv(0,t)), 
\end{equation}
where $\partial_2$ is the derivative with respect to the second component, then expanding the left-hand-side of Eq.~\eqref{SlowMap} as a Taylor series about $t = nT$ using the definition of $\xv_{n+1}$ from Eq.~\eqref{xns} gives that 
\begin{equation}\label{Duality}
    \Fv(\xv) = \xv + \varepsilon T \Gv(\xv) + \mathcal{O}(\varepsilon^2).
\end{equation}
Hence, when $\varepsilon$ is small Eq.~\eqref{Duality} represents an Euler step of the slow physics. This demonstrates the duality between the functions $\Fv$ and $\Gv$, at least to $\mathcal{O}(\varepsilon^2)$. Therefore, discovering the mapping $\Fv$ from data presents a novel method to extract the slow timescale physics of a given signal, something which holds great potential for multiscale systems that are so common in the biological and medical sciences \cite{Alber}.

\begin{figure}[t] %Figure: Overview of the method
%\center
\includegraphics[width=0.5\textwidth]{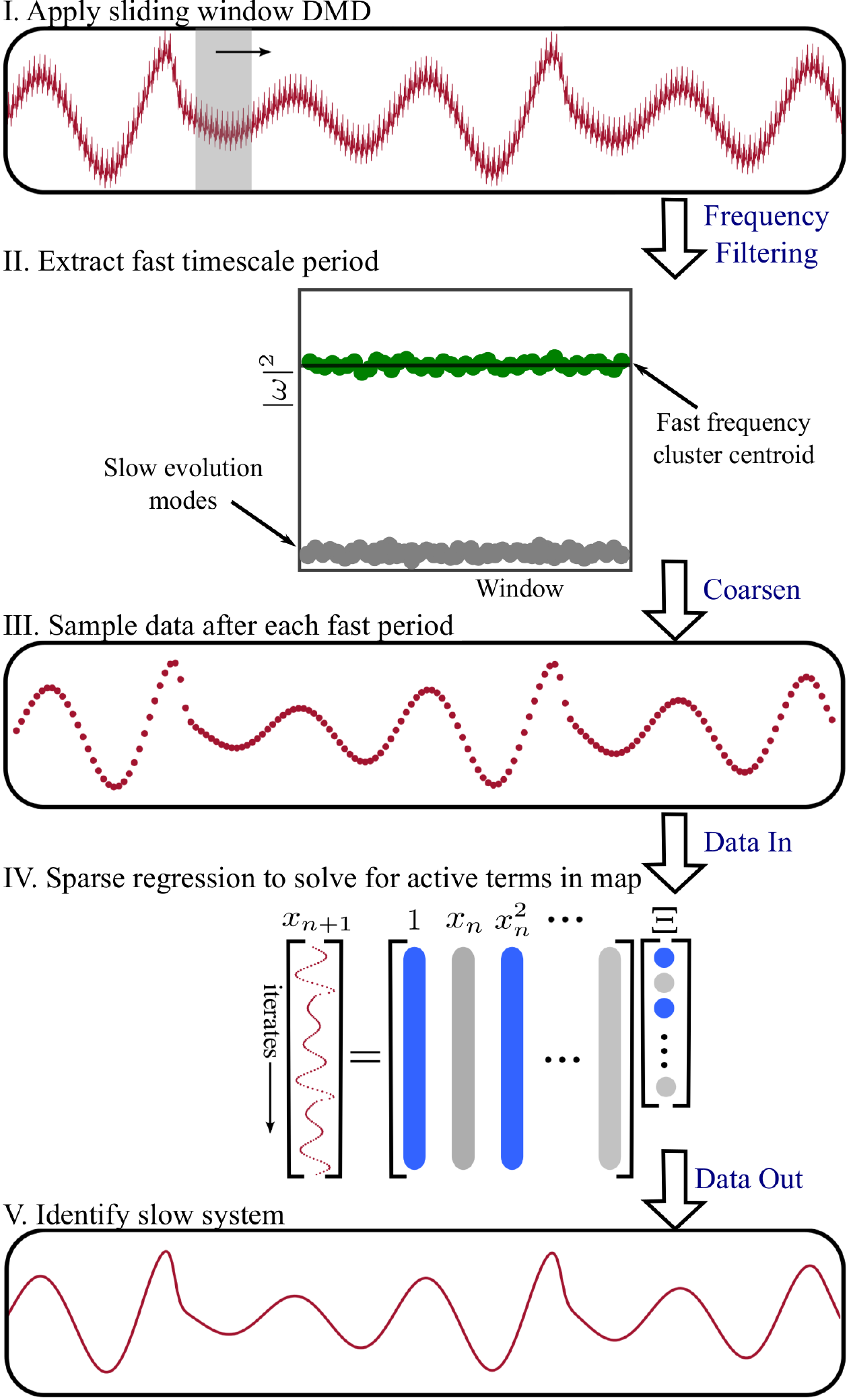}
\caption{An overview of the method presented in this work. We first use the sliding window DMD technique \cite{Dylewsky} to identify the fast period, then track the signal at integer multiples of the fast period, resulting in the coarsened signal $x_n = \Phiv(0,\varepsilon nT)$. We then identify a mapping $x_n \mapsto x_{n+1}$ using the SINDy method \cite{SINDy}, and finally use the expansion Eq.~\eqref{Duality} to extract the slow timescale physics.}
\label{fig:Slow_Extraction}
\end{figure}

There are potentially two unknowns that are required to find $\Gv$: the fast period $T>0$ and the mapping $\Fv$. Here we use the method of sliding-window {\em dynamic mode decomposition} (DMD)~\cite{Dylewsky} to extract the fast period and the {\em sparse identification of nonlinear dynamics} (SINDy) algorithm \cite{Bramburger,SINDy} to obtain the mapping $\Fv$. We summarize the method visually in Figure~\ref{fig:Slow_Extraction} and explain the individual components in more detail in the following section. The advantage our proposed method presents is that it does not attempt to discover a dynamical system for the multiscale signal, since as we show in Section~\ref{sec:Multiscale}, sparse identification procedures should be expected to fail at this task in general. This is due in part to the numerical approximation of the derivative of the signal, which in the presence of a periodic fast timescale is rapidly changing. Although the challenge of approximating the derivative can be overcome by using instead integral terms \cite{Reinbold,Schaeffer}, the present method circumvents this difficulty altogether. Furthermore, the multiscale property of the data presents a unique advantage in the implementation of a regression technique which promotes sparsity using thresholding: unsupervised timescale separation is expected to introduce error at $\mathcal{O}(\varepsilon^2)$, whereas the sparsification procedure amounts to pruning terms of order $\mathcal{O}(\varepsilon)$. As a result, the method by construction recovers only the leading order slow timescale dynamics. 

The sparse regression framework has been used previously to discover multiscale physics~\cite{Champion}, but with a much different procedure for the model discovery process and sampling strategy.  Specifically, fast sampling of the dynamics was required since both fast- and slow-timescale physics needed to be discovered, whereas in this work, the fast scale is known and only the slow dynamics needs to be discovered.  Modification to the basic SINDy \cite{SINDy} architecture allows for the discovery of partial differential equations~\cite{Rudy}, equations of rational expressions~\cite{Kaheman}, and conservation laws~\cite{Kaiser}.  For each of these modifications, one can imagine using the proposed method to handle multiscale temporal dynamics in an efficient manner. 

This paper is organized as follows. In Section~\ref{sec:Methods} we present a detailed overview of the method for extracting slow timescale physics. This includes summaries of the two major components of the method: the sliding window DMD technique and the SINDy algorithm. In Section~\ref{sec:Multiscale} we provide a simple example of a multiscale system where naive application of the SINDy method fails to properly identify the full governing equations. We further supplement this discussion by showing that our methods accurately capture the slow evolution of this system once the fast dynamics have been taken out of the signal. Section~\ref{sec:Applications} is entirely dedicated to applications of the method, where we present examples with discovered slow dynamics that are monotone, periodic, and chaotic. Finally, we briefly summarize our findings in Section~\ref{sec:Conclusion}.

\section{\label{sec:Methods} Methods}

The method employed by this work has two major components: scale separation using the sliding-window DMD and sparse regression analysis using SINDy. In this section we briefly summarize these methods and direct the reader to the works \cite{Dylewsky} and \cite{SINDy}, respectively, for a more complete discussion of each component. In its totality, our method first employs the sliding window DMD technique to identify the fast period, then track the signal at integer multiples of the fast period, resulting in the coarsened signal $\xv_n = \Phiv(0,\varepsilon nT)$. We then identify a mapping $\Fv$ such that $\xv_{n+1} = \Fv(\xv_n)$ using the SINDy method. Finally, we use the duality Eq.~\eqref{Duality} to extract the slow timescale physics when $\varepsilon$ is sufficiently small. We summarize the method visually in Figure~\ref{fig:Slow_Extraction}, and proceed through the following subsections with a discussion of the individual components of our method.

\subsection{\label{subsec:DMD}Scale separation using sliding-window DMD}

The method presented in this work relies on foreknowledge of the timescale disparity between components constituting a multiscale signal. For systems with known governing equations, this can generally be determined by inspection of the coefficients or by perturbation expansion. For systems with unknown equations of motion, however, a data-driven discovery method must suffice. This could be accomplished by identifying peaks on the Fourier spectrum, for example. For multiscale physics, windowed Fourier transforms can help provide improved resolution of signals and their content.  Indeed, such windowing procedures are the basis of multi-resolution analysis and wavelet decompositions~\cite{mallat}.  Multi-resolution DMD~\cite{mrdmd} provides a analogous decomposition of multivariate data, identifying coherent spatial modes and temporal frequencies in multiscale systems.

Dynamic Mode Decomposition (DMD) is a model regression technique which seeks a best-fit linear representation for observed dynamics~\cite{schmid,dmdbook}. Given a data matrix $\Xv$ consisting of $m$ sequential snapshots, DMD identifies a linear operator $\boldsymbol{A}$ which in some optimal (least-squares) sense satisfies the equation $\dot{\Xv} = \boldsymbol{A}\Xv$. This approximation is of course unlikely to be accurate for highly nonlinear systems, but this problem can be circumvented by subsampling $\Xv$ onto shorter time intervals by sliding a window of width $d$ ($d \ll m$) across the full time series. Even when global dynamics are nonlinear, local linear approximations are easily obtainable. Each local operator $\boldsymbol{A}$ has eigenvalues which characterize the timescale content of local dynamics.

In this work we use a technique introduced by Dylewsky et al. (2019) which leverages DMD to separate the components of multiscale data and learn local linear models for the dynamics by clustering on their eigenspectra \cite{Dylewsky}. This method has the advantages of sparsity, flexibility, and robustness to overlapping or highly disparate dynamical timescales. By gathering eigenspectra across all sliding-window iterations of DMD, one can form a statistical picture of the global spectral content of the data.  A simple clustering algorithm can identify the most prominently represented timescales and offer a parsimonious estimate of the scale components present. Even when timescales are separated by many orders of magnitude, the algorithm can be applied recursively with varied window width to properly identify them. These clusters can be depicted visually by plotting the modulus squared of the eigenvalues of the local operators $\boldsymbol{A}$ for each window. In the present scenario of a signal satisfying Eq.~\eqref{Signal} at least two distinct clusters should be apparent: one bounded away from 0 representing the fast periodic dynamics and another near 0 representing the slow dynamics (which should appear nearly static over the comparatively short span of a windowed subsample). A cartoon of this clustering is presented in panel II of Figure~\ref{fig:Slow_Extraction}, while an application of the method to real data is presented in Figure~\ref{fig:Toy_Cluster}.

\subsection{\label{SINDy}Sparse identification of nonlinear dynamics (SINDy)}

The SINDy method, introduced by Brunton et al. (2016), is an algorithm for discovery of a symbolic representation of the governing equations of a system from time series measurements~\cite{SINDy}. Given a data matrix comprised of sequential state measurement snapshots:
\begin{equation}
    \Xv_1 = \left[
                    \begin{matrix}
                        \text{ ---} & \xv^T(t_1) &  \text{--- }\\
                        \text{ ---} & \xv^T(t_2) &  \text{--- }\\
                         & \vdots &  \\
                        \text{ ---} & \xv^T(t_m) &  \text{--- }
                    \end{matrix}\right]
\end{equation}
along with another data matrix of the same size, $\Xv_2$, comprised of either the temporal derivative of the data at the same measurement times or the successive iterates of the discrete temporal data. A symbolic representation for the dynamics $\dot{\xv}(t) = \boldsymbol{F}(\xv(t))$ or $\xv(t_{n+1}) = \Fv(\xv(t_n))$ is constructed from a library of candidate functions for terms of $\Fv(\xv)$. The chosen functions are evaluated on the measurement data to construct a library matrix $\Theta$ whose $m$ rows represent the $m$ measurement snapshots of $\Xv_1$ lifted into a space of all library observables. For example, a library consisting of polynomials up to degree 2 for $\xv = (x_1,x_2)\in\mathbb{R}^2$ would look like
\begin{equation}
    \begin{split}
    &\Theta(\Xv_1) = \\ 
    &\begin{bmatrix}
        1 & x_1(t_1) & x_2(t_1) & (x_1(t_1))^2 & x_1(t_1)x_2(t_1) & (x_2(t_1))^2\\
        1 & x_1(t_2) & x_2(t_2) & (x_1(t_2))^2 & x_1(t_2)x_2(t_2) & (x_2(t_2))^2\\
        \vdots & \vdots & \vdots & \vdots & \vdots & \vdots\\
        1 & x_1(t_m) & x_2(t_m) & (x_1(t_m))^2 & x_1(t_m)x_2(t_m) & (x_2(t_m))^2\\
      \end{bmatrix}
    \end{split}
\end{equation}
The claim that the equation of motion $\Fv(\xv)$ is some linear combination of the chosen library functions is equivalent to the statement
\begin{equation}\label{eq:sindy_regression}
    \Xv_2 = \Theta(\Xv_1)\Xi \quad 
\end{equation}
for some coefficient matrix $\Xi$. For many systems of interest the governing equations contain only a few terms, so a sparseness requirement can be imposed on $\Xi$. This can be carried out using any sparse regression algorithm on the over-determined system of linear equations (\ref{eq:sindy_regression}), e.g. elastic net methods such as LASSO or Ridge. The approach employed in this paper, as in \cite{SINDy}, is a sequentially-thresholded least squares method in which Eq.~\eqref{eq:sindy_regression} is initially solved simply by minimizing $\|\Xv_2 - \Theta(\Xv_1)\Xi\|_2$. Elements of $\Xi$ with magnitude smaller than some chosen sparsity parameter $\lambda > 0$ are then thresholded to $0$, and the process repeats. This iteration is carried out until all remaining coefficients have absolute value larger than $\lambda$. This process is a proxy for $\ell_0$ optimization~\cite{Zheng}. It has convergence guarantees~\cite{Zhang} and performs comparably to LASSO in most cases at significantly reduced computational expense. Moreover, the direct correspondence between the sparsity parameter $\lambda$ and the obtained dynamical coefficients in $\Xi$ proves useful in the multiscale case treated in this paper: we show that knowledge of the timescale components present in a signal can be used to determine {\em a priori} a suitable value for $\lambda$.

We note that there are many variants of sparse regression, all of which typically attempt to approximate a solution to an $NP$-hard, $\ell_0$-norm penalized regression.  Sparsity-promoting methods like the LASSO~\cite{tibshirani1996regression,su2017false} use the $\ell_1$-norm as a proxy for sparsity since tractable computations can be performed.   The iterative least-squares thresholding algorithm of the SINDy algorithm promotes sparsity through a sequential procedure.  Recently, Zhang and Schaeffer~\cite{Zhang} have provided a number of rigorous theoretical results on the behavior and convergence of the SINDy algorithm. Specifically, they prove that the algorithm approximates local minimizers of an unconstrained $\ell_0$-penalized least-squares problem.  This allows them to provide sufficient conditions for general convergence, the rate of convergence, and conditions for one-step recovery. As shown in Champion et al.~\cite{champion2019unified}, the SINDy regression framework does not readily accommodate extensions, additional constraints, or improvements in performance.  Thus the optimization formulation is extended to include additional structure, robustness to outliers, and nonlinear parameter estimation using the {\em sparse relaxed regularized regression} (SR3) approach that uses relaxation and partial minimization~\cite{Zheng}.  Rigorous theoretical bounds are provided for the relaxed formulation which has three advantages over the state-of-the-art sparsity-promoting algorithms: 1) solutions of the relaxed problem are superior with respect to errors, false positives, and conditioning; 2) relaxation allows extremely fast algorithms for both convex and nonconvex formulations; and 3) the methods apply to composite regularizers, essential for total variation (TV) as well as sparsity-promoting formulations using tight frames. Indeed, the SR3 formulation was shown to have superior performance (computational efficiency, higher accuracy, faster convergence rates, and greater flexibility) across a range of regularized regression problems with synthetic and real data, including applications in compressed sensing, LASSO, matrix completion, TV regularization, and group sparsity. Between the three recent papers~\cite{champion2019unified,Zheng,Zhang}, rigorous estimates can be established for this model-discovery paradigm.

\section{\label{sec:Multiscale}SINDy and multiscale systems}

\begin{figure} %Figure: Quasiperiodic solution x_1(t)
%\center
\includegraphics[width=0.45\textwidth]{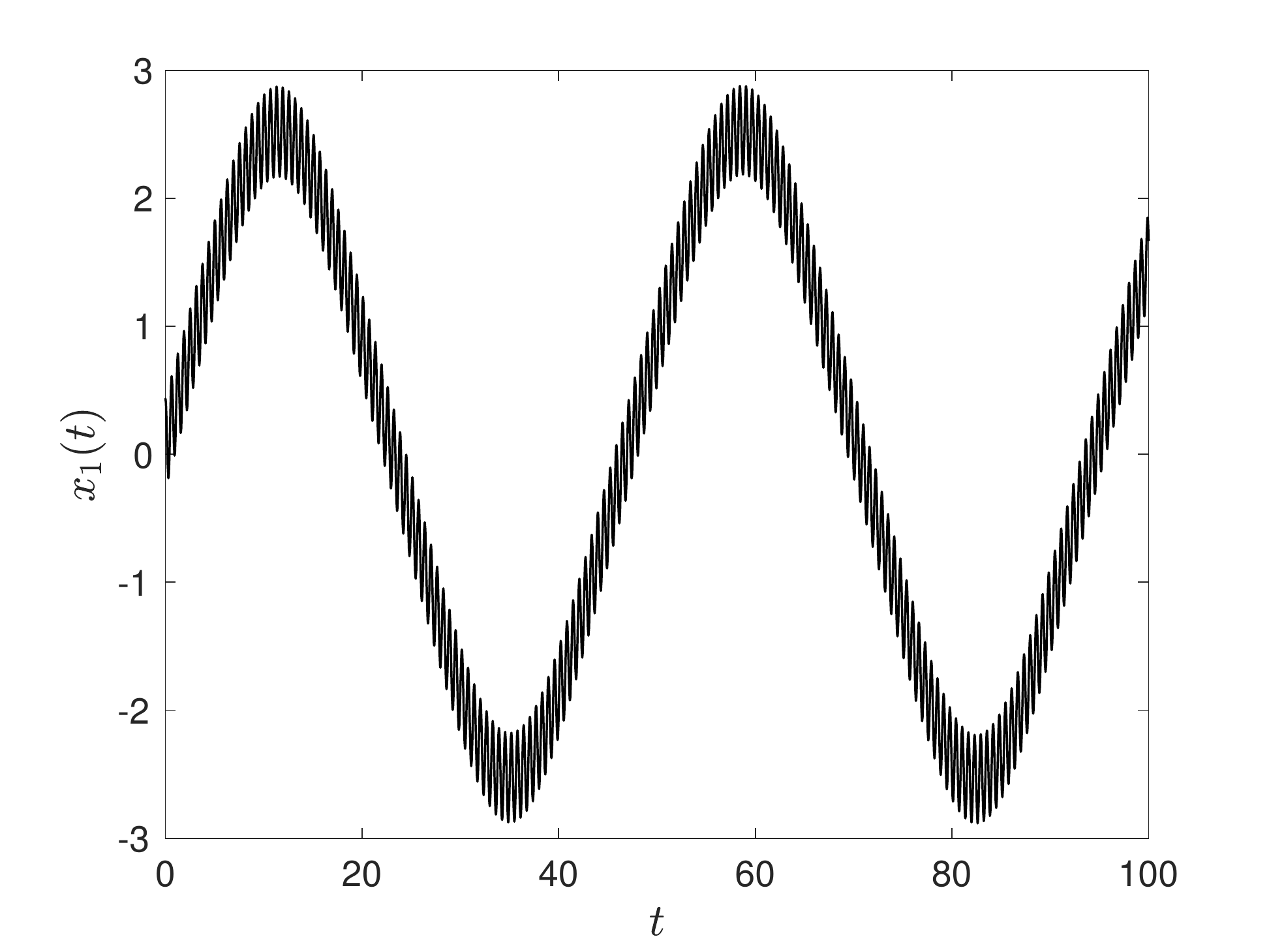}
\caption{The first component of the signal $\xv(t)$ in Eq.~\eqref{ToySignal}. There are two distinct timescales and the signal is plotted for slightly more than two periods of the slow timescale. The fast period is approximately $\pi/5$.}
\label{fig:QuasiPeriodic}
\end{figure}

In this section we briefly demonstrate how naively applying the SINDy method to a signal in the form of \eqref{Signal} to discover both the fast and slow timescales simultaneously should be expected to fail, thus necessitating the method described in this manuscript.  Alternatively, one can us fast sampling strategies to resolve the discovery process~\cite{Champion}, but here sampling on the slow scale is all that is needed.  In~\cite{Dylewsky} a simple toy model was created to extract the periods of oscillation for signals formed as linear combinations of periodic phenomena. The model is given by
\begin{equation}\label{ToyModel} %Toy model
	\begin{split}
		\dot{v}_1 &= v_2, \\
		\dot{v}_2 &= -w_1^2v_1^3, \\
		\dot{w}_1 &= w_2, \\
		\dot{w}_2 &= -100w_1 - 4w_1^3,
	\end{split}
\end{equation}
where constants are chosen to appropriately separate the timescales. The $(w_1,w_2)$ variables are governed by an unforced Duffing equation, for which almost all initial conditions fall into steady periodic motion. The $(v_1,v_2)$ variables form a cubic oscillator with a coefficient $w_1^2$ dependent on the state of $w_1$. The signal \eqref{Signal} is produced by integrating system~\eqref{ToyModel} using MATLAB's ode23 function with a maximal timestep of $10^{-4}$ and initial conditions $(v_1,v_2,w_1,w_2) = (0,0.5,0,0.5)$. We then take a randomly generated orthogonal matrix $Q \in \mathbb{R}^{4\times 4}$ to define
\begin{equation}\label{ToySignal}
	\xv(t) = Q\cdot[v_1(t),v_2(t),w_1(t),w_2(t)]^T,
\end{equation}
in an effort to sufficiently mix the disparate temporal dynamics. We direct the reader to Figure~\ref{fig:QuasiPeriodic} for a visual depiction of the first component of $\xv(t) = (x_1(t),x_2(t),x_3(t),x_4(t))$, and note that the rest look similar in that we can clearly see the two timescales present in the signal $\xv(t)$. 

\begin{figure} %Figure: SINDy Reconstruction
\center
\includegraphics[width=0.45\textwidth]{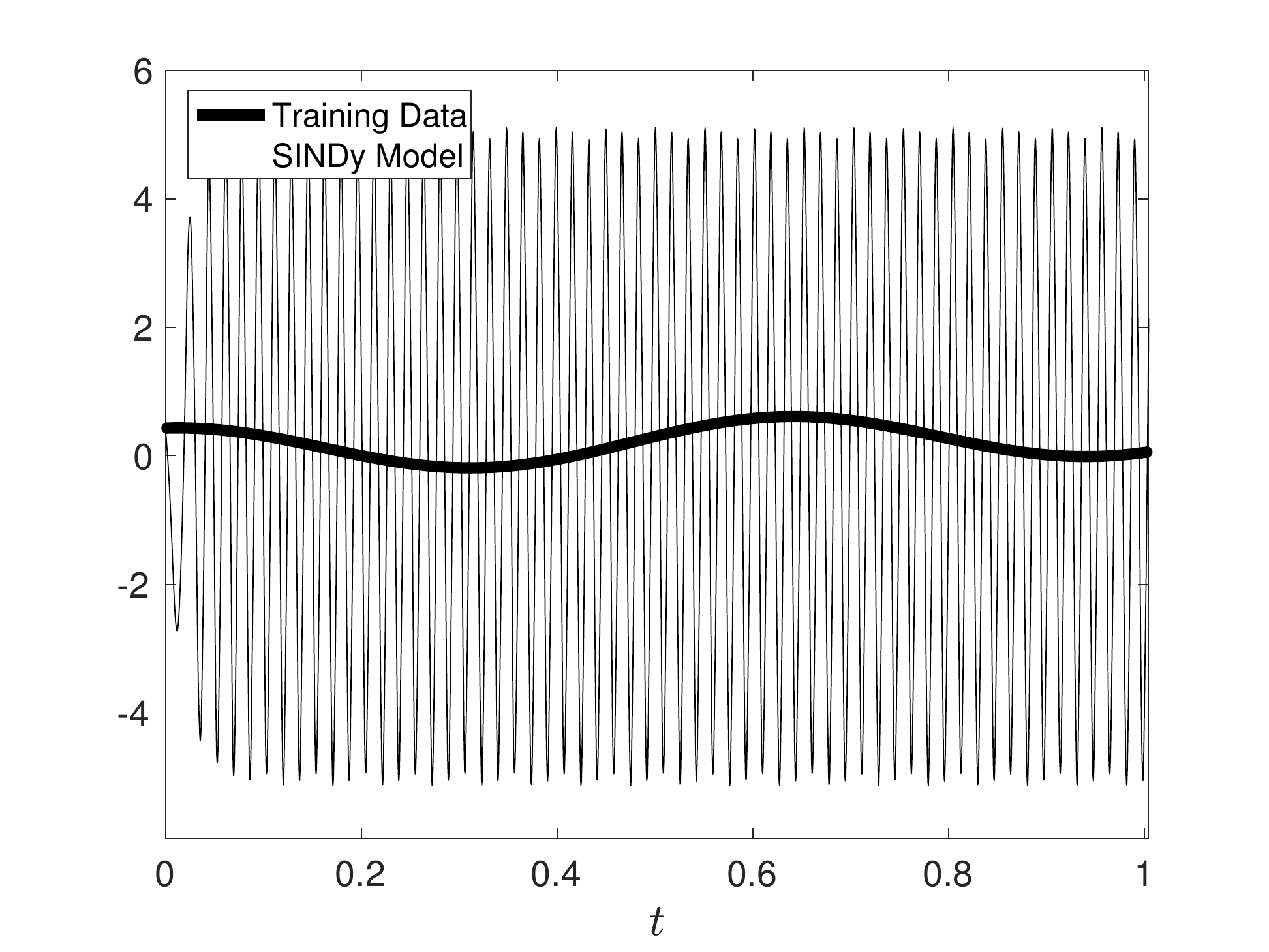}
\caption{A comparison of the first component of the training data against the evolution of the discovered continuous-time SINDy model with the same initial data. Notice that the discovered data has a significantly higher frequency of oscillation, as can be observed by comparing the extent of the horizontal axis in this figure with that of Figure~\ref{fig:QuasiPeriodic}.}
\label{fig:Toy_Comparison}
\end{figure}

If we use the SINDy method for discovery of a continuous-time dynamical systems using Eq.~\eqref{ToySignal} as the training data and a sparsity parameter $\lambda = 10^{-3}$, we discover the differential equation
\begin{equation}\label{Toy_Discovery}
	\begin{split}
		\dot{x}_1 &= 0.1 - 788.8x_1 + 788.0x_4  + 3.3x_4^3, \\
		\dot{x}_2 &= -833.3x_1 + 833.3x_4, \\
		\dot{x}_3 &= -0.1 - 877.8x_1  + 878.7x_4 - 3.3x_4^3, \\
		\dot{x}_4 &= -0.3 - 922.3x_1  + 924.0x_4 - 6.6x_4^3.
	\end{split}
\end{equation}  
Even without knowing the exact values in $Q$ we immediately note that this discovered system cannot exactly capture the dynamics of the signal since it includes constant terms. In Figure~\ref{fig:Toy_Comparison} we plot the first component of the training data against the dynamics of the discovered model \eqref{Toy_Discovery} with the same initial condition. We can see that the discovered model produces a signal with a faster frequency of oscillation than that of the original signal. This can be observed by comparing the extent of the horizontal axes of Figures~\ref{fig:QuasiPeriodic} and \ref{fig:Toy_Comparison}, where the former displays the signal for $t \in [0,100]$ while the latter uses $t \in [0,1]$. This significantly shorter time window in Figure~\ref{fig:Toy_Comparison} is meant to emphasize the extremely fast frequency of oscillation in the discovered continuous-time SINDy model, so it should be noted that the slow dynamics of Eq.~\eqref{ToySignal} are are not yet visible on such a short timescale. We further note that changing the sparsity parameter $\lambda$ has little effect on the discovered model, always producing a dynamical system which fails to even approximately reproduce the dynamics of the input signal \eqref{ToySignal}. The repository {\bf GitHub/jbramburger/Slow-Discovery} contains all code related to this example to ensure the ability to reproduce these results and perform further experiments.

This example is merely meant to illustrate what a naive application of the continuous-time SINDy method can result in when applied to multiscale training data. We suspect that the discrepancy between the SINDy model and the training data is due at least in part to the approximation of the derivative. The fast frequency of oscillation present in the multiscale training data results in a rapidly oscillating derivative, which requires finely spaced data. Hence, one way to improve the performance of the SINDy model might be to take smaller timesteps in the training data, but we note that decreasing the maximal step size down a full order of magnitude to $10^{-5}$ shows little improvement in the discovered continuous-time SINDy model. Of course, in the case of real data it seems unlikely that such a re-generation of the data could be performed and is therefore not a practical method for improving the performance of the discovery algorithm. Another method would be to avoid taking numerical derivatives altogether and instead use weak formulations which replace the derivative with integral terms \cite{Reinbold,Schaeffer}. The strength of the method presented herein is that we are able to consider coarse-grained data which experiences little variation from one data point to the next (representing the slow dynamics), thus avoiding this difficulty altogether.

\begin{figure} %Figure: SINDy Reconstruction
\center
\includegraphics[width=0.45\textwidth]{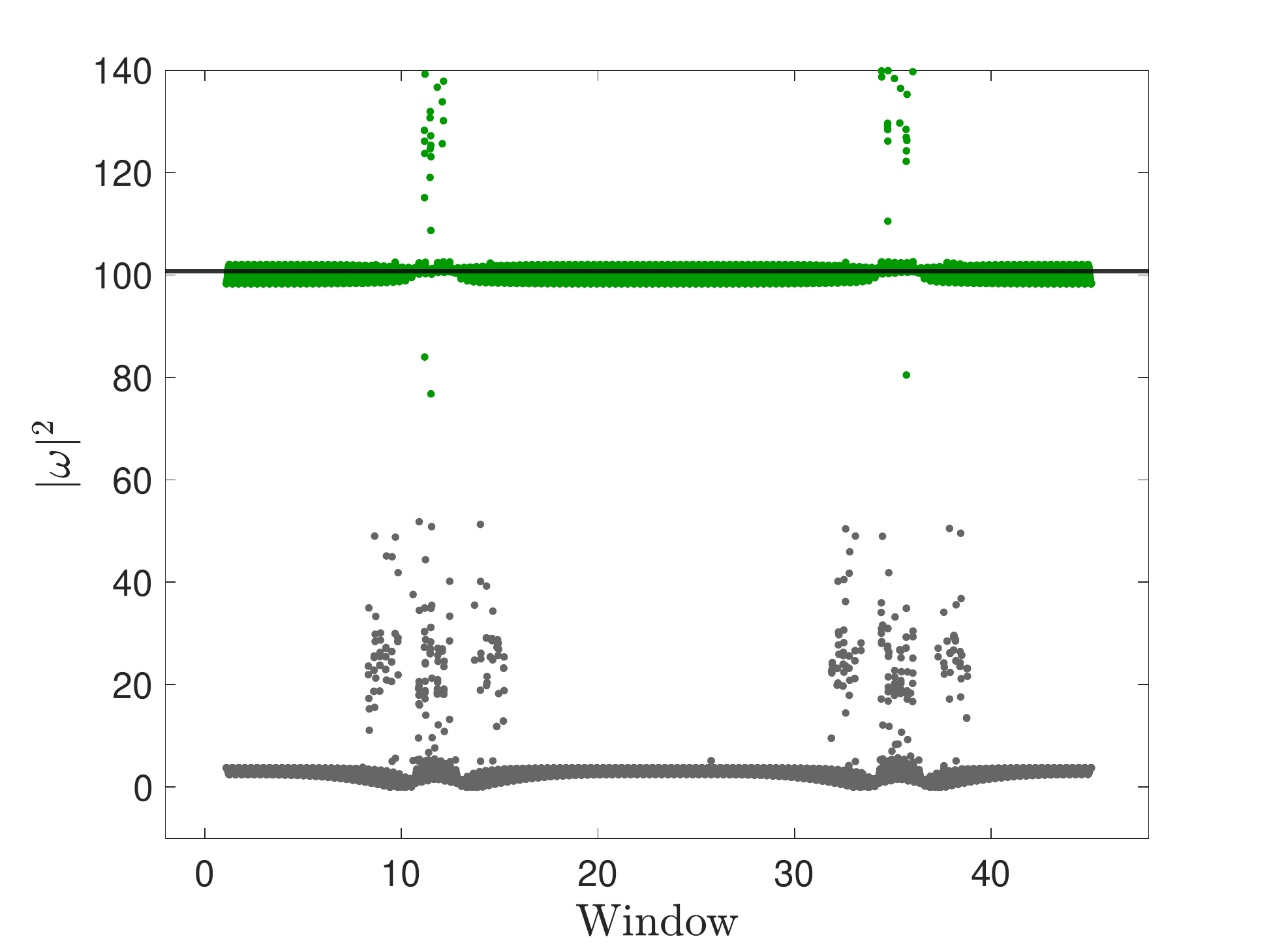}
\caption{The result of the sliding window DMD method applied to the signal~\eqref{ToySignal}. The horizontal axis represents the value of $t$ at which the window is centred, while the vertical axis plots the square modulus of the purely imaginary eigenvalues $\omega$ of the local operator {\bf A} for each window. Colours are used to delineate between the two clusters, representing the fast and slow components of the signal. The horizontal black line represents the centroid of the fast frequency cluster, approximately $|\omega|^2 = 100$.}
\label{fig:Toy_Cluster}
\end{figure}

We now turn to coarsening the signal by tracking it at integer multiples of the fast timescale period. The training data in this case is exactly the same as that used to discovery the continuous-time SINDy model \eqref{Toy_Discovery}. In Figure~\ref{fig:Toy_Cluster} we plot the resulting frequencies $\omega$ extracted from the sliding window DMD method, where one can clearly see the separation between fast and slow dynamics. The centroid of the fast cluster gives that the fast component of the signal has period given by $T = \pi/5$, and hence we can track the full signal at integer multiples of this fast period. We comment that the frequencies showing the largest variation from the clusters are localized to windows near the extreme points of the slow dynamics, representing regions where the slow dynamics are not approximately constant in the relatively small windows. It is exactly this problem that necessitates the sliding window method in the first place, as looking at the whole signal will inevitably introduce large discrepancies since the slow dynamics show large variation on long timescales. 

\begin{figure} %Figure: Quasiperiodic mapping results
\includegraphics[width=0.45\textwidth]{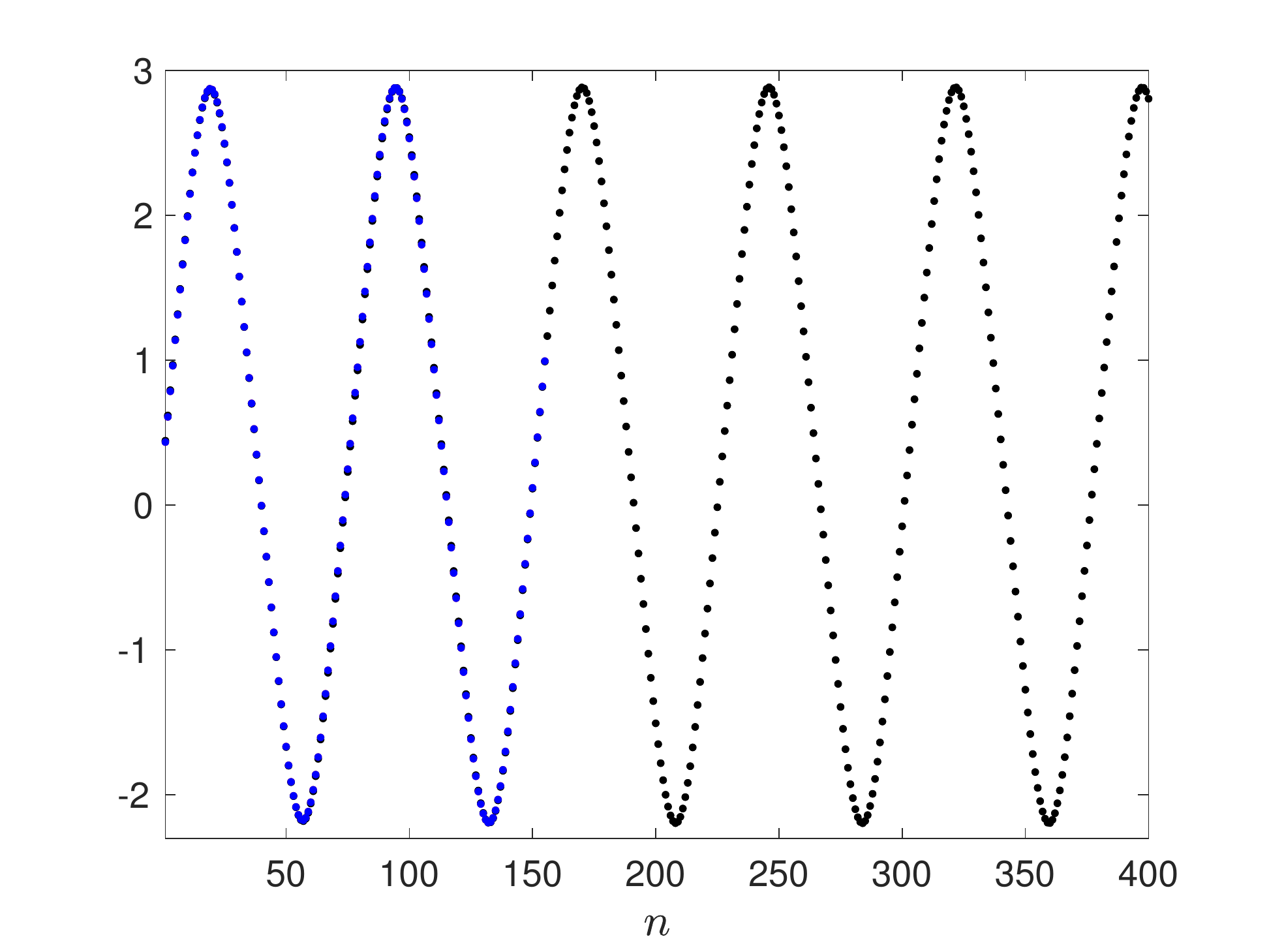}
\caption{Iterates of the discovered mapping for the slow timescale dynamics of the first component of \eqref{ToySignal}, seeded with the initial value of the training data. Training data is provided in blue along with the iterates of the discovered mapping in black.} %Training data uses approximately 2 full periods of the slow scale dynamics amounting to slightly more than 150 periods of the fast scale dynamics.}
\label{fig:QuasiPeriodic_Map}
\end{figure}

Having now obtained the fast period of oscillation, we are able to apply the sparse identification procedure to obtain a mapping over the slow dynamics. The training data for the SINDy method uses 200 fast periods - long enough to observe at least two full slow periods. The iterates of the component $x_1(nT)$ to the discovered mapping $\Fv$ are presented in Figure~\ref{fig:QuasiPeriodic_Map} against the training data, while we note that the other components look similar. The cumulative error over all components is 1\%, a significant improvement from the model \eqref{Toy_Discovery}. Hence, we see that in the absence of the fast oscillations the SINDy method is able to effectively track the slow evolution of the signal. 

\section{\label{sec:Applications}Applications}

Let us consider a simple and motivating example. Consider the scalar ODE
\begin{equation}\label{Logistic}
	\dot{x} = \varepsilon x(1 - x + \sin(2\pi t)), \quad x(0) = x_0 \in\mathbb{R}_+
\end{equation}
with $0 < \varepsilon \ll 1$. Since the ODE is $1$-periodic in the independent variable $t$, the theory of averaging for dynamical systems \cite{Guckenheimer,Sanders} dictates that there exists a constant $C > 0$ such that the solution $x(t)$ satisfies $|x(t) - y(t)| \leq C\varepsilon$ for all $t \geq 0$ \footnote{Technically, the theory dictates that this bound only holds for $t$ on a timescale of length $1/\varepsilon$, but since all positive initial conditions of Eq.~\eqref{Logistic} evolve towards a global attractor, the bound can be shown to hold for all $t \geq 0$ so long as the bound $C > 0$ depends on $x_0$.}, where $y(t)$ is a solution of the autonomous ODE
\begin{equation}\label{AveragedLogistic}
	\dot{y} = \varepsilon y (1-y), \quad y(0) = x_0.
\end{equation}
The solution $y(t)$ is explicitly given by
\[
	y(t) = \frac{x_0}{x_0 + (1 - x_0)\exp(-\varepsilon t)},
\] 
where we can therefore see that $y(t)$ evolves on a slow timescale, $\varepsilon t$. Hence, $y(t)$ makes up the slow dynamics to at least $\mathcal{O}(\varepsilon)$ and using Eq.~\eqref{Duality} we expect the mapping \eqref{SlowMap} to be of the form $\Fv(x) = x + \varepsilon x(1-x) + \mathcal{O}(\varepsilon^2)$. 

To illustrate the performance of our method, we take $\varepsilon = 10^{-2}$ and a sparsity parameter $\lambda = \varepsilon^2$ to truncate at order $\varepsilon^2$. We discover the mapping \eqref{SlowMap} here to be
\begin{equation}\label{SINDyLogistic}
    \begin{split}
	\Fv(x) &= 1.001x - 0.009778x^2 \\
	&\approx x + \varepsilon x(1-x) + \mathcal{O}(\varepsilon^2),	
    \end{split}
\end{equation}
conforming with our expectation from the above analysis. Furthermore, the leading order terms in Eq.~\eqref{SINDyLogistic} represent a forward Euler discretization of Eq.~\eqref{AveragedLogistic} with step size $\varepsilon$, implying that standard error bounding arguments based on the initial condition and the value of $\varepsilon$ can be applied to bound the difference between the solutions of \eqref{Logistic} at integer values of $n\geq 0$ and the iterates of the map \eqref{SINDyLogistic}. 

This example illustrates how the method performs against a benchmark ODE where the slow dynamics can be explicitly determined via analysis. In reality, few systems which exhibit multiscale dynamics take the form of singularly perturbed dynamical systems and therefore even determining the fast timescale period to average over is a nontrivial task.

\subsection{\label{subsec:Planets}Planetary dynamics}

As in the toy model example of Section~\ref{sec:Multiscale}, multiscale phenomena can often be observed when there are two periodic components of the signal with a vast separation between their periods. That is, consider a signal which can approximately be written as 
\begin{equation}\label{DoublePeriodic}
	\xv(t) = P_0(t) + P_1(t),
\end{equation}
where $P_0$ is periodic with period $T_0> 0$ and $P_1$ is periodic with period $T_1 > 0$. Assuming that $T_0 \ll T_1$, naturally leads to the scale separation parameter 
\[
	\varepsilon := \frac{T_0}{T_1} \ll 1.
\]
We may rescale $t = T_0\tau$ so that $\tilde{P}_0(\tau) := P_0(T_0\tau)$ is now $1$-periodic and $P_1(T_0\tau)$ is $\varepsilon^{-1}$-periodic. Setting 
\[
	\tilde{P}_1(s) := P_1(\varepsilon^{-1}s) 
\]  
makes $\tilde{P}_1$ $1$-periodic as well. Hence, the full signal $x(\tau)$ is can equivalently be written
\[
	\tilde{\xv}(\tau) := \xv(T_0\tau) = \tilde{P}_0(\tau) + \tilde{P}_1(\varepsilon \tau)
\]
for which both functions $\tilde{P}_0$ and $\tilde{P}_1$ are $1$-periodic and clearly demonstrate the separation of timescales of interest in this work. The mapping \eqref{SlowMap} tracks $\tilde{\xv}(n)$ for integers $n \geq 0$, or equivalently, $\xv(nT_0)$.  

Such a phenomenon was exemplified in the model~\eqref{ToyModel} and can further be found in the motion of the planets Saturn and Jupiter around the sun, as described by a three-body planetary model. For this test case we generate data by constructing a system of three objects subject to pairwise gravitational attraction (using the non-relativistic Newtonian formulation). The masses and initial conditions of these objects are assigned based on measurements of the true Jupiter-Saturn-Sun system in its center-of-mass reference frame. Time series trajectories in $\mathbb{R}^3$ are obtained via symplectic numerical integration over a period of $10^6$ years.

\begin{figure} %Figure: Jupiter and Saturn orbits
\center
\includegraphics[width=0.5\textwidth]{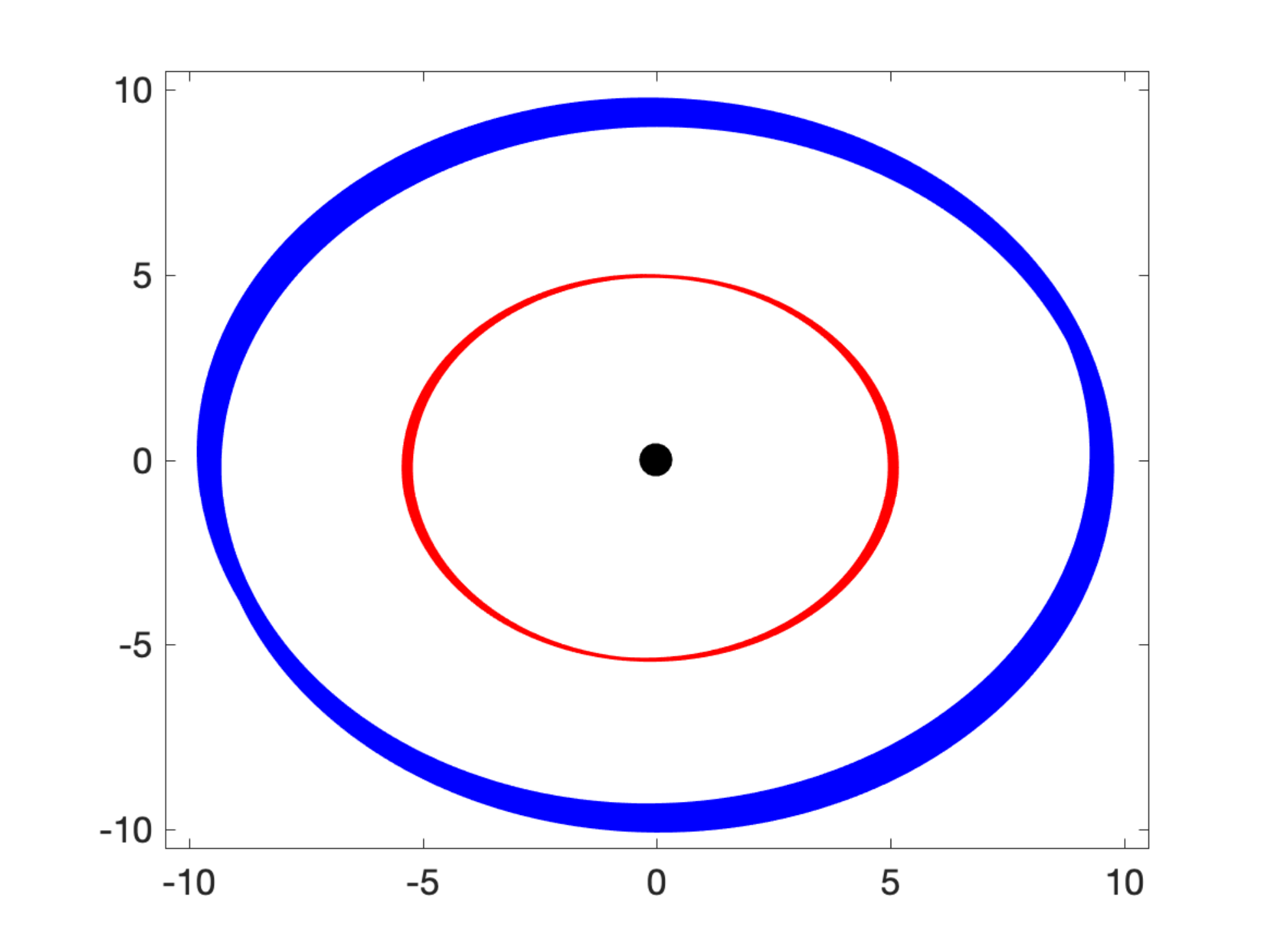}
\caption{The procession of Jupiter (red) and Saturn (blue) about the sun (black dot) in their orbital plane in nondimensionalized coordinates. The fast timescale for each planet constitutes simple procession about the sun: 11.86 years for Jupiter and 28.5 years for Saturn. Beyond this fast procession are a number of slow scales that drive the eccentricity of the orbit, giving the appearance that they trace out thick circles in the orbital plane over thousands of years.}
\label{fig:Orbits}
\end{figure}

The signal for each planet contains multiple different timescales, and the motion of each planet in its orbital plane can be approximated by a signal of the form~\eqref{DoublePeriodic}. The periodic function $P_0$ describes the primary orbit of the planets around the sun, for which upon applying the sliding window DMD technique we find periods $T_0 = 11.86$ years for Jupiter and $T_0 = 29.5$ years for Saturn. The function $P_1$ describes the eccentricity of these orbits, for which the sliding-window DMD technique has shown that $T_1 \approx 46800$ years~\footnote{It was shown in \cite{Dylewsky} that Jupiter has a periodic component with period between our $T_0$ and $T_1$, but this period is almost exactly $9T_0$ and is expected to be a numerical artifact.} for both planets \cite{Dylewsky}. This periodic eccentricity of the orbits comes from the interaction of the two planets orbiting the relatively massive sun and can be observed in Figure~\ref{fig:Orbits} where we plot their orbits around the sun in their orbital plane.  

Let us begin by considering our signal to be the motion of Jupiter projected entirely into its orbital plane over time. Here we have $T_0 = 11.86$ with $t$ measured in years and so from the construction above we get $\varepsilon \approx 2.6\times 10^{-4}$. Denote $x^J_n$ and $y^J_n$ to be the $x$ and $y$ components of the position of Jupiter in its orbital plane at time $t = nT_0 = 11.86n$ years. Taking a sparsity parameter $\lambda$ on the order of $\varepsilon^2 \sim 10^{-8}$ is impractical since numerical error present in either simulating the data or implementing the SINDy method should be expected to show up in the discovered mapping. To overcome this, we note that $T_0\varepsilon \sim 10^{-3}$, and so properly discovering $\Gv$ per Eq.~\eqref{Duality} is expected to succeed with $ 10^{-3} \leq \lambda < T_0\varepsilon$ since the lower bound should be large enough to eliminate numerical error and $\mathcal{O}(\varepsilon^2)$ terms, while the upper bound allows for discovery of the $\mathcal{O}(\varepsilon)$ terms. With $\lambda = 10^{-3}$ we discover the slow-scale mapping 
\begin{equation}\label{JupiterForecast}
	\begin{bmatrix}
		x^J_{n+1} \\ y^J_{n+1}	
	\end{bmatrix} = \begin{bmatrix}
		0.0036 \\ -0.0018
	\end{bmatrix} + \begin{bmatrix}
		0.9999 & 0.0114 \\ -0.0114 & 0.9999
	\end{bmatrix} \cdot \begin{bmatrix}
		x^J_{n} \\ y^J_{n}	
	\end{bmatrix} 
\end{equation}
using training data comprised of 1500 fast timescale periods - long enough to observe at least two full periods of the first slow timescale. The mapping \eqref{JupiterForecast} can be used to find that the slow timescale mapping $\Gv$ in this case is given by
\begin{equation}\label{JupiterSlow}
    \Gv(x,y) = \begin{bmatrix}
		0 & 1 \\ -1 & 0
	\end{bmatrix}\cdot\begin{bmatrix}
		1.17 + 3.69x \\ 0.57 + 3.69y
	\end{bmatrix}
\end{equation}
using Eq.~\eqref{Duality}. Hence, we see that the eccentricity dynamics of Jupiter is given, to leading order, by an ellipse in the orbital plane. We note that the conservative structure of the slow dynamics is not necessarily guaranteed by the conservative structure of the original three-body problem since it could be the case that energy flows from one scale to the next. Despite this, the resulting conservative structure of Eq.~\eqref{JupiterSlow} could potentially reflect that the training data is stable up to $\mathcal{O}(\varepsilon^2)$ perturbations on very long timescales. Hence, we do not expect that Eq.~\eqref{JupiterSlow} is valid for all time, but only long finite timescales until evolution on even slower timescales begins to influence the $\mathcal{O}(\varepsilon)$ dynamics of Jupiter's orbit.  

We may proceed in the same way for the procession of Saturn around the sun to find a similar result. In this case we have $T_0 = 29.5$ years, making up the fast scale period, and therefore the scale separation parameter is given by $\varepsilon \approx 6.3\times 10^{-4} $. Performing our mapping discovery procedure again with $\lambda = 10^{-3}$, for the same reasons as in the case of Jupiter, results in the mapping  
\begin{equation}\label{SaturnForecast}
	\begin{bmatrix}
		x^S_{n+1} \\ y^S_{n+1}	
	\end{bmatrix} = \begin{bmatrix}
		-0.0038 \\ 0.0025
	\end{bmatrix} + \begin{bmatrix}
		1.0000 & -0.0100 \\ 0.0100 & 1.0001
	\end{bmatrix} \cdot \begin{bmatrix}
		x^S_{n} \\ y^S_{n}	
	\end{bmatrix} 
\end{equation}
where the variables $(x^S_n,y^S_n)$ represent the orthogonal components of the position of Saturn in its orbital plane at $t = 29.5n$ years. We can similarly rearrange Eq.~\eqref{SaturnForecast} via Eq.~\eqref{Duality} to find that the eccentricity dynamics of Saturn are confined to an ellipse in the orbital plane.

\subsection{\label{subsec:Chaos}Chaotic slow dynamics}

\begin{figure*} %Figure: Slow scale Chaos
%\center
\includegraphics[width=\textwidth]{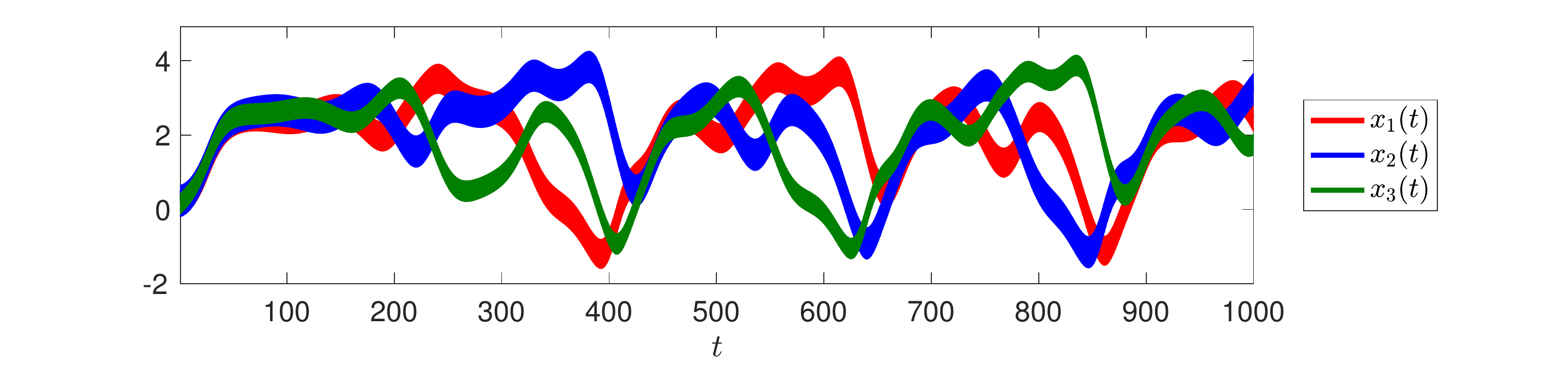} \\
\includegraphics[width=\textwidth]{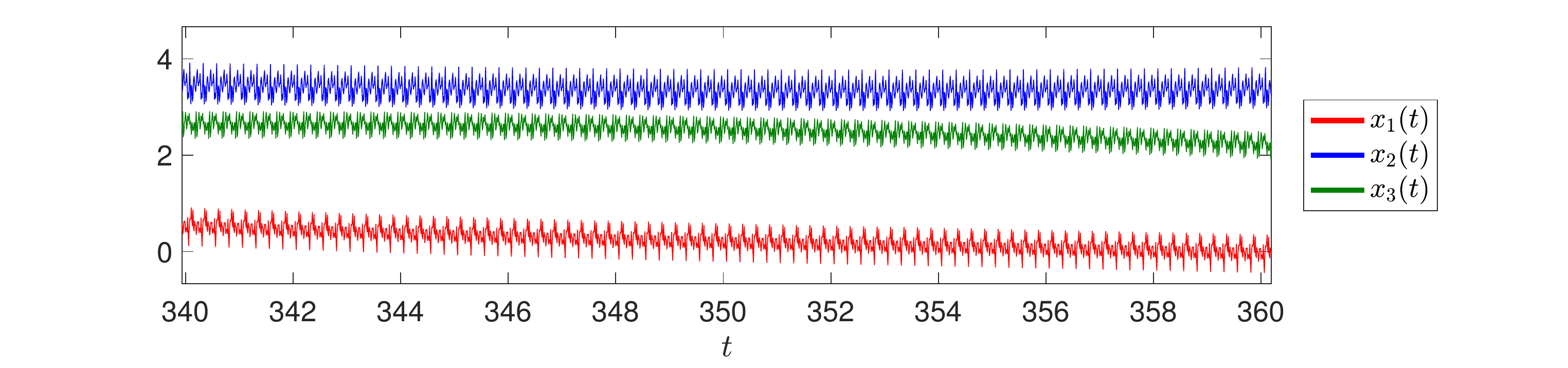} 
\caption{A signal of the form \eqref{ChaosSignal} with periodic fast dynamics given by Eq.~\eqref{PerChaos} with randomly-generated Fourier coefficients and slow dynamics governed by the system~\eqref{Chaos}. The top shows the signal over a long timescale so that the chaotic dynamics can be observed. The bottom shows a smaller time window so that only the periodic fast scale can be observed.}
\label{fig:Chaos}
\end{figure*}

In this example we apply the method to systems for which the slow timescale dynamics are chaotic. We consider a signal 
\begin{equation}\label{ChaosSignal}
	\xv(t) = \varepsilon P(t) + C(\varepsilon t),
\end{equation}
where $P(t)$ is periodic with period $T > 0$ and $C(t)$ is a trajectory of a chaotic dynamical system. For the purpose of illustration we will take $C(t)$ to be a trajectory on the circularly symmetric attractor of Thomas~\cite{Thomas}, given by the three-dimensional dynamical system
\begin{equation}\label{Chaos}
	\begin{split}
		\dot{C}_1 &= \sin(C_2) - 0.2C_1, \\
		\dot{C}_2 &= \sin(C_3) - 0.2C_2, \\
		\dot{C}_3 &= \sin(C_1) - 0.2C_3,
	\end{split}
\end{equation}
with initial condition $C(0) = (0.3,0.2,0.1)^T$. The damping value $0.2$ is chosen so that the system is indeed chaotic and the initial condition was chosen arbitrarily to produce a trajectory on the chaotic attractor. The periodic signal $P(t)$ will take the form of a truncated Fourier series 
\begin{equation}\label{PerChaos}
	P(t) = a_0 + \sum_{n = 1}^N \bigg[a_n\cos\bigg(\frac{2\pi t}{T}\bigg) + b_n\sin\bigg(\frac{2\pi t}{T}\bigg)\bigg]
\end{equation}
for some $N > 1$ fixed and coefficients $a_0,a_n,b_n\in\mathbb{R}^3$. For our work here we will fix $N = 10$ and note that working with larger or smaller $N$ produces nearly identical results. We further take the coefficients $a_n$ and $b_n$ to be drawn from the uniform distribution on the unit cube $[-1,1]^3$, fix $\varepsilon = 0.1$ and $T = \frac{1}{4}$, which efficiently separates the timescales. The reader is referred to Figure~\ref{fig:Chaos} for a characteristic illustration of the signal.

As before we will fix $\lambda = \varepsilon^2 = 0.01$ to ensure that only $\mathcal{O}(\varepsilon)$ terms are present in the discovered equation. However, since the discovered equation contains sinusoidal functions, a library containing only monomials will never be able to fully reproduce the dynamics of the slow scale. Hence, the best one could hope for is to produce a monomial series representation of the dynamics of \eqref{Chaos}, but the truncation at $\mathcal{O}(\varepsilon^2)$ will inevitably truncate the series representation as well. Furthermore, the chaotic nature of system~\eqref{Chaos} leads one to conjecture that SINDy will not necessarily just return a truncated Taylor series representation for $\sin(x)$ since the elements of $C(t)$ do not remain small on large timescales. This is confirmed by implementing the discovery method 100 different times, resulting in 100 different randomized functions $P(t)$, and discovering $\Fv(\xv) = \xv + \mathcal{O}(\varepsilon^2)$ in every case, meaning that the slow timescale dynamics are unable to be picked up at all with such a monomial basis. This emphasizes that the choice of basis functions plays potentially an even more critical role in the case of discovering the dynamics of the slow timescale system than it potentially would for discovering systems with a single timescale.   

The inclusion of both sine and cosine terms into the library immediately remedies the above problem. For example, one implementation of this method resulted in the system
\begin{equation}\label{DiscoverChaos}
	\Fv(\xv) = \begin{bmatrix}
		0.02508\sin(x_2) + 0.9949x_1  \\
		0.02428\sin(x_3) + 0.9947x_2  \\
		0.02471\sin(x_1) + 0.9951x_3
	\end{bmatrix},
\end{equation}
which is characteristic of all implementations with randomized Fourier coefficients in Eq.~\eqref{PerChaos}. After rearranging, the slow dynamical system $G(\xv)$ is recovered as
\[
    \Gv(\xv) = \begin{bmatrix}
	    1.0032\sin(x_2) - 0.2040x_1  \\
		0.9712\sin(x_3) - 0.2120x_2 \\
		0.9884\sin(x_1) - 0.1960x_3
	\end{bmatrix},
\]
which agrees with Eq.~\eqref{Chaos} up to $\mathcal{O}(\varepsilon^2)$. \vspace{.1 in}

\section{\label{sec:Conclusion}Conclusion}

In this work we have seen a computationally cheap and efficient method for discovering the slow timescale dynamical system of a multiscale signal that is amenable to averaging. This method relies on tracking the signal not continuously, but after each fast period, to produce a mapping which parsimoniously describes the slow physics. The resulting mapping can also be related to the continuous-time dynamics of the slow timescale via the Euler stepping scheme Eq.~\eqref{Duality}. Furthermore, the fast period can be extracted accurately using the recently developed sliding window DMD technique with clustering of eigenfrequencies \cite{Dylewsky}, which can also be used to reconstruct the fast component of the signal. The result is an algorithm that reliably reproduces the slow evolution of a multiscale signal using DMD for the fast timescale and SINDy for the slow timescale. Given the difficulty in approximating the emergent slow scale evolution dynamics of multiscale systems, the method provides a viable architecture for coarse-graining to achieve accurate, interpretable and parsimonious dynamical models for slow-scale physics. Finally, we note that depending on the scaling, averaging can yield trivial slow dynamics, in which case homogenization is the correct procedure \cite{Givon}. This potentially limits the class of signals for which our techniques are applicable, although our work herein provides an initial step towards data-driven discovery of the slow timescale dynamics of multiscale signals.  

\begin{acknowledgments}
JJB was supported by a PIMS PDF held at the University of Victoria. JNK acknowledges support from the Air Force Office of Scientific Research (AFOSR) grant FA9550-17-1-0329.
\end{acknowledgments}


\begin{thebibliography}{99} %References

    \bibitem{Alber} M. Alber, A.B. Tepole, W.R. Cannon, S. De, S. Dura-Bernal, K. Garikipati, G. Karniadakis, W.W. Lytton, P. Perdikaris, L. Petzold, and E. Kuh. Integrating machine learning and multiscale modeling - perspectives, challenges, and opportunities in the biological, biomedical, and behavioral sciences, {\em npj Digit. Med.} {\bf 2}, (2019) 115.

    \bibitem{andersen} H. C. Andersen. Molecular dynamics simulations at constant pressure and/or temperature. {\em J. Chem. Phys.} {\bf 72}, (1980) 2384-2393.

	\bibitem{Bramburger} J.J. Bramburger and J.N. Kutz. Poincar\'e maps for multiscale physics discovery and nonlinear Floquet theory, {\em Physica D} {\bf 408}, (2020) 132497.
	
	\bibitem{Brasser} R. Brasser, A.C. Barr, and V. Dobos. The tidal parameters of TRAPPIST-1b and c, {\em Mon. Not. R. Astron. Soc.} {\bf 487}, (2019) 34–47.
	
	\bibitem{SINDy} S.L. Brunton, J.L. Proctor, and J.N. Kutz. Discovering governing equations from data by sparse identification of nonlinear dynamical systems, {\em P. Natl. Acad. Sci. USA} {\bf 113}, (2016) 3932-3937.	
	
	\bibitem{Champion} K.P. Champion, S.L. Brunton, and J.N. Kutz. Discovery of nonlinear multiscale systems: Sampling strategies and embeddings, {\em SIAM J. Appl. Dyn. Syst.} {\bf 18}, (2018) 312-333. 
	
	\bibitem{champion2019unified} K. Champion, P. Zheng, A. Aravkin, S.L. Brunton, and J.N. Kutz. A unified sparse optimization framework to learn parsimonious physics-informed models from data, (2019) arXiv:1906.10612.
	
	\bibitem{Dylewsky} D. Dylewsky, M. Tao, and J.N. Kutz. Dynamic mode decomposition for multiscale nonlinear physics, {\em Phys. Rev. E} {\bf 99}, (2019) 063311.

	\bibitem{Givon} D. Givon, R. Kupferman, and A. Stuart. Extracting macroscopic dynamics: Model problems and algorithsm, {\em Nonlinearity} {\bf 17}, (2004) R55-R127.

	\bibitem{Guckenheimer} J. Guckenheimer and P. Holmes. {\em Nonlinear oscillations, dynamical systems and bifurcations of vector fields}, Springer-Verlag, New York, (1983). 
	
	\bibitem{Kaheman} K. Kaheman, J.N. Kutz, S.L. Brunton. SINDy-PI: A robust algorithm for parallel implicit sparse identification of nonlinear dynamics, arXiv:2004.02322.
	
	\bibitem{Kaiser} E. Kaiser, J.N. Kutz, S.L. Brunton. Discovering conservation laws from data for control, in: {\em 2018 IEEE Conference on Decision and Control, CDC}, (2018) 6415–6421.
	
	\bibitem{karplus} M. Karplus and J. A. McCammon.  Molecular dynamics simulations of biomolecules, {\em Nat. Struc. Bio.} {\bf 9}, (2002) 646-652.
	
	\bibitem{Kifer} Y. Kifer. Averaging principle for fully coupled dynamical systems and large deviations, {\em Ergod. Theor. Dyn. Syst.} {\bf 24}, (2004) 847-871.
	
	\bibitem{dmdbook} J. N. Kutz, S. L. Brunton, B. W. Brunton, B.W. and J. L. Proctor. {\em Dynamic mode decomposition: data-driven modeling of complex systems}. Society for Industrial and Applied Mathematics (2016).
	
	\bibitem{mrdmd} J. N. Kutz, X. Fu, and S. L. Brunton. Multiresolution dynamic mode decomposition. {\em SIAM J. Appl. Dyn. Syst.} {\bf 15}, (2016) 713-735.
	
	\bibitem{Laskar} J. Laskar. Secular evolution of the solar system over 10 million, {\em Astron. Astrophys.} {\bf 198}, (1988) 341.
	
	\bibitem{Malhotra} R. Malhotra, M. Holman, and T. Ito. Chaos and stability of the solar system, {\em P. Natl. Acad. Sci. USA} {\bf 98}, (2001) 12342-12343.
	
	\bibitem{mallat} S. Mallat. {\em  A wavelet tour of signal processing}. Elsevier, (1999).

	\bibitem{Palus} M. Palus. Multiscale atmospheric dynamics: Cross-frequency phase-amplitude coupling in the air temperature, {\em Phys. Rev. Lett.} {\bf 112}, (2014) 078702.
	
	\bibitem{Reinbold} P.A.K. Reinbold, D.R. Gurevich, and R.O. Grigoriev. Using noisy or incomplete data to discover models of spatiotemporal dynamics, {\em Phys. Rev. E} {\bf 101}, (2020) 010203.
	
	\bibitem{Rudy} S.H. Rudy, S.L. Brunton, J.L. Proctor, and J.N. Kutz. Data-driven discovery of partial differential equations, {\em Sci. Adv.} {\bf 3}, (2017) e1602614.
	
	\bibitem{Sanders} J.A. Sanders, F. Verhulst, and J. Murdock. {\em Averaging methods in nonlinear dynamical systems}, Springer-Verlad, New York, (2007).
	
	\bibitem{Schaeffer} H. Shaeffer and S. McCalla. Sparse model selection via integral terms, {\em Phys. Rev. E} {\bf 96}, (2017) 023302.
	
	\bibitem{schmid} P. J. Schmid. Dynamic mode decomposition of numerical and experimental data. {\em J. Fluid Mech.} {\bf 656}, (2010) 5-28.
	
	\bibitem{su2017false} W. Su, B. Ma, and E. Candes. False discoveries occur early on the lasso path, {\em Ann. Stat.} {\bf 45}, (2017) 2133-2150.
	
	\bibitem{tibshirani1996regression} R. Tibshirani. Regression shrinkage and selection via the lass, {\em J. R. Stat. Soc.} {\bf 58}, (1996) 267-288.
	
	\bibitem{Thomas} R. Thomas. Deterministic chaos seen in terms of feedback circuits: Analysis, synthesis, `labyrinth chaos', {\em Int. J. Bifurc. Chaos} {\bf 9}, (1999) 1889-1905.	

    \bibitem{Wunsch} C. Wunsch. The long-period tides, {\em Rev. Geophys.} {\bf 5}, (1967) 447-475.

    \bibitem{Wunsch2} C. Wunsch, D.B. Haidvogel, and M. Iskandarani. Dynamics of the long-period times, {\em Prog. Oceanogr.} {\bf 40}, (1997) 81-108.
    
    \bibitem{Zhang} L. Zhang and H. Schaeffer. On the convergence of the SINDy algorithm, {\em Multiscale Model. Sim.} {\bf 17}, (2019) 948-972.
    
    \bibitem{Zheng} P. Zheng, T. Askham, S.L. Brunton, J.N. Kutz, and A.Y. Aravkin. A unified framework for sparse relaxed regularized regression: Sr3, {\em IEEE Access} {\bf 7}, (2019) 1404-1423.
    
\end{thebibliography}
\end{document}